# Shooting Over or Under the Mark:
# Towards a Reliable and Flexible Anticipation in the Economy

Koichiro Matsuno
Department of BioEngineering
Nagaoka University of Technology
Nagaoka 940-2188, Japan

**Abstract**
The real monetary economy is grounded upon monetary flow equilibration or the activity of actualizing monetary flow continuity at each economic agent except for the central bank. Every update of monetary flow continuity at each agent constantly causes monetary flow equilibration at the neighborhood agents. Every monetary flow equilibration as the activity of shooting the mark identified as monetary flow continuity turns out to be off the mark, and constantly generate the similar activities in sequence. Monetary flow equilibration ceaselessly reverberating in the economy performs two functions. One is to seek an organization on its own, and the other is to perturb the ongoing organization. Monetary flow equilibration as the agency of seeking and perturbing its organization also serves as a means of predicting its behavior. The likely organizational behavior could be the one that remains most robust against monetary flow equilibration as an agency of applying perturbations.

## 1 Introduction

Once the idea of equilibrium between the supplier and the consumer of a commodity is employed for an economic theorization, the movement leading to an equilibrium between the two is strictly distinguished from the movement displacing the equilibrium point. Consider, for instance, the play at a tug of war. The equilibrium theory distinguishes between the standstill movement of each player putting the whole energy into tugging and the displacement movement of all of the participating players in one direction in the end. However, the distinction does not tell anything significant to the participating players because each player is just tugging the rope with the fullest strength. Even if a spectator says that the tug of war consists of two different kinds of movement, this observation would remain most irrelevant to the participating players. To be sure, the equilibrium theory is excellent in precipitating a standstill equilibrium among the competing agencies. It must be a greatest achievement in economic theorization. Nonetheless, there is also the other side of the achievement. The real economy to be addressed is not so elegant and simplistic as the equilibrium theory would imply. What has to be focused at this point is to see how the movement that may seem to displace the equilibrium point could come to be actually realized.

The economic dynamics going beyond the stipulation of the equilibrium theory can be seen more concretely in the national monetary economy (Matsuno, 1978a). Any nation state has the sovereignty of letting its central bank have the monopoly of issuing the bank notes. In the nation state monetary economy, every economic agent except for the central bank is required to observe the continuity of monetary flow between those coming into and going out. The condition of monetary flow continuity has to be observed in the record compiling the completed transactions, otherwise either unlawful counterfeit or destroying of the bank notes must have been committed. At the least, the monetary authority keeps watching and recording the continuity of monetary flow registered in the present perfect tense. The financial statistics issued regularly by the monetary authority is a sort of the balance sheet on the state scale. One can find in the balance sheet a whole array of the accounting identities supplemented and specified by each concrete figure including the outstanding of the bank notes issued by the central bank. On the other hand, the monetary economy is a system in movement, which is descriptively in the present progressive tense. There thus arises a distinct contrast between the record in the present perfect and the movement in the present progressive tense. This remark now lets us face a serious endeavor of figuring out how one could salvage the movement in progress from the vast record registered in the present perfect tense.

In particular, theoretical endeavor of elucidating movement in progress requires a stressful attention. If a movement in progress is adequately described in the present tense, it would remain legitimate at any present moment and then turn out to be equivalent to something remaining in stasis. The movement that the equilibrium theory brings about is just this sort of movement equivalent to being in stasis. If it is intended to uphold the movement in the present progressive tense upon the observation made in the present tense, both movement and stasis would have to be one and the same. Precisely for this reason, any movement going beyond the stipulation set by the equilibrium theory has to ground itself upon the observation made in other than the present tense. It is required to reach the present progressive tense from other than the present tense. One candidate satisfying this requirement which this article is going to explore in the following is to reach the present progressive tense from the present perfect. That is to say, it will be attempted to retrieve movement in progress from the record registered in the present perfect tense.

**2 Formulation of the Problem**

The incontrovertible ground upon which the movement of the monetary economy is based is monetary flow continuity in the record registered in the present perfect tense. The continuity observed in the perfected record is however no more than the consequence of the effort on the part of participating agents. What is more, the preceding consequence of fulfilling the condition of monetary flow continuity at one agent constantly serves as a cause for inducing the similar activities of actualizing monetary flow continuity at other agents in the neighborhood (Matsuno, 1978a).

Consider, for instance, the case that the government increases the tax rate in order to finance the deficit in its revenue. The effect of temporary recovery of monetary flow continuity at the government must now induce such a reverberation that households may decrease their consumption. Corporations may also decrease their capital investment to restore monetary flow continuity at each of them. The aftereffect of the decrease in both consumption and capital investment must further urge the government to revise its previous decision so as to recover monetary flow continuity by, say, issuing the governmental bonds more. There should be no end in the reverberating effort for recovering monetary flow continuity. The government may even increase the tax rate further in order to finance the unexpected deficit on the road to the inevitable next updating. Monetary flow equilibration as the activity of fulfilling monetary flow continuity at each economic agent constantly spills over into the economy.

Constant reverberation for monetary flow equilibration is imputed to transactions that are intrinsically bilateral. No bilateral transactions can fully take into account the whole array of transactions proceeding concurrently in the entire economy. It is only after the events when any transaction partner can get acquainted with other transactions concurrently in progress. Every update of monetary flow continuity at each economic agent comes to be disturbed subsequently when the reverberation imputed to the similar concurrent updates in the neighborhood reaches. Monetary flow equilibration necessarily intended to shoot the mark identified as monetary flow continuity turns out to fail as being too short or too far off the mark every time. Every economic transaction is constantly swaying between overshooting and undershooting the mark. A typical example is the haphazard ups and downs of the tax rate or the amount of securities issued by the government.

It is however not the exception but the rule to observe that every effort for shooting the mark eventually comes to fail in meeting the challenge while the effort itself tirelessly survives in any part of the economy. This autonomous activity of shooting the mark while ending up with its failure every time underlies both organizing and perturbing the economy from within. The monetary economy is thus seen as an organization transforming the organizational activity in progress constantly into disorganizing perturbations in the effect.

**3 Syntactic Integration and Semantic Concretization**

The monetary economy as an organizing body drives itself from within. Every activity of monetary flow equilibration is organizational in the making, but perturbing in the effect. The apparently contradictory aspect of monetary flow equilibration between organizing and perturbing is however the reality of the monetary economy. It is of course desirable to figure out a way of mitigating the apparent contradiction between the two modes of behavior. One attempt in this direction must be to take notice of the categorical difference between the two modes of organizing and perturbing (Matsuno, 2000).

The activity of fulfilling the condition of monetary flow continuity is integrative or synthetic for the sake of the economic organization to be maintained. Still, what is implied is abstract and not concrete enough to specify the activity. Monetary flow equilibration as an organizing activity in progress is syntactic in the abstract domain in the sense that a syntactic integration is aimed at from within while there is no specification in advance. In contrast, monetary flow equilibration in the effect is already concrete enough to specify its semantic implication in the empirical domain in that it goes beyond the syntactic stipulation and then relates itself to the activities in the empirical monetary economy. What is going to perturb the economic organization is just this sort of semantic concretization.

Every activity in the monetary economy is concrete enough, and has no need of referring to abstract notions. Nonetheless, once it is intended to descriptively gain access to the de facto organization, some descriptive means for the purpose must necessarily be invoked. That is syntactic integration in the abstract domain. Despite its limitation as a linguistic artifact, syntactic integration is viewed as a descriptively inevitable means of organizing the train of semantic concretizations, the latter of which alone does not tell anything specific before the events. Syntactic integration certainly serves as an organizing principle referring to the situations prior to the events thanks to its abstract nature.

The interplay between syntactic integration and semantic concretization, when applied to the actual monetary economy, reveals that the ongoing organization is constantly perturbed by its own deed. Perturbations consequential upon the preceding organization induces inconsistencies in the semantics in the sense that there have arisen concrete events mutually conflicting in the empirical domain because of the absence of any means to coordinating everything to every other globally on the spot. What should be significant at this point is that semantic inconsistencies may be tolerable in the empirical domain to the extent that syntactic integration in the abstract domain can be kept intact. This is because syntax by definition cannot tolerate any inconsistency internally. Semantic inconsistencies now serve as both the cause and the effect of syntactic organization. One thus recognizes that the most likely syntactic organization is the one that could remain most robust against semantic inconsistencies. Those organizations vulnerable to semantic inconsistencies would most likely lose the chance of their realization (Matsuno, 1989). In other words, the most likely syntactic organization is the one that can take most advantage of semantic inconsistencies for the sake of its own sustenance.

**4 Retrieving the Present Progressive Tense**
Syntactic organization feeding on semantic inconsistencies, if it can survive for some time, would leave its own record registered in the present perfect tense. Such a surviving organization can be anchored at the irrevocable record while its movement in progress is in the present progressive tense. This contrast comes to focus how to relate the present perfect to the present progressive tense in the operation of syntactic organization while constantly feeding on semantic inconsistencies generated from within. More specifically, it is required to retrieve the present progressive tense from the record registered in the present perfect since any empirical science including economics is based upon unequivocal records of empirical data more than anything else.

The issue will be how to retrieve the activity of shooting over or under the mark identified as monetary flow continuity from the record already satisfying the stipulation of monetary flow continuity at each economic agent. At first sight, this problem itself would seem ill-suited because it asks us to retrieve causal activities from the accounting identities already satisfying monetary flow continuity. No progressive movement can be envisioned within the accounting identities, since the latter of which are consequences of the causal activities rather than the other way around. The accounting identities in the perfect tense definitely include in themselves inevitable artifacts preventing the occurrence of movement in the present progressive tense. Despite that, movement in progress by no means disappears from the actual monetary economy.

Imagine, for instance, that the amount of both cash and demand deposit listed in the asset in the balance sheet filed by a corporation at a certain term end is more than enough. This may mean that the management would already have committed to either paying back the debt or making a new investment, or an arbitrary mix of the two alternatives. Otherwise, the management would be blamed as being incompetent by the stockholders. Conversely, if the amount of both cash and demand deposit is too short to meet the required payment, the management would commit the crime called breach of trust towards the stockholders.

The management always runs the risk of something like walking on the very narrow fence. If fallen on either side, the management would necessarily be blamed. The actual movement in progress the management exhibits is demonstrated in the activity of fulfilling the accounting identities known as monetary flow continuity so that the consequential balance sheet may not be blamed in whatever respects. The balance sheet itself does not demonstrate any of such movement in progress.

The preceding effect of shooting the mark identified as monetary flow continuity does necessarily fail subsequently, with the consequence of inducing the similar activities of shooting ad infinitum. Shooting either over or under the mark in the effect is the rule in every effort of shooting the mark. This observation may invite us to admit that a surviving organization of the monetary economy would be hard to be envisioned, but not quite so in reality. One can be more optimistic. This haphazardness can in turn give the monetary economy its robustness and stability against perturbations from within. The most robust organization, that is necessarily syntactic in its makeup and most likely to occur, is the one that can take most advantage of the then available perturbations. The present perspective may help us even to figure out how the record registered in the present perfect tense could be related to the movement in the present progressive tense.

The global record registered in the present perfect tense is, however, an artifact in the best sense of the word, whether it may represent the balance sheet on the national scale compiled by the monetary authority from time to time regularly. The record always assumes a specific external observer, such as the monetary authority, whose job is to compile the record while prohibiting itself from participating in the actual transactions upholding the recorded monetary economy at the same time. One of the inevitable interferences from the presence of the external observer compiling the record is the forced synchronization in the data editing. The record assumes the presence of a global clock to be referred to by every datum. All of the figures appeared in the balance sheet strictly and unanimously refer to the same end of the term when the report was complied. Data in the record are globally synchronized in the sense that the transactions are globally synchronized. In contrast, the actual transactions are undoubtedly bilateral and the accompanied synchronization is at most bilateral. Although data in the record are compiled as observing a global synchronization, the actual transactions in progress proceed at most in bilateral synchronization. This distinction between global and bilateral synchronization now provides us with a framework of properly perceiving the three of syntactic integration, semantic concretization and semantic inconsistencies underlying the organization of the monetary economy.

First of all, the record registered in the present perfect tense is an embodiment of semantic concretization because it is about a linguistic representation going beyond the abstract notion intrinsic to the syntactic rule. Semantic concretization in the present perfect tense is however an artifact in the respect of assuming a global synchronization. Once the stipulation of global synchronization is lifted for the sake of the actual bilateral one, the record in the eyes of the participating economic agents can be seen concrete enough but as demonstrating inconsistencies among themselves. Although they are forcibly concealed by semantic concretization assuming a global synchronization, semantic inconsistencies definitely survive in bilateral synchronization. It is the present semantic inconsistencies that are responsible for driving the syntactic organization of the monetary economy.

Our problem is a bit convoluted. The driving factor for the organization is semantic inconsistencies, while what have been available in the record is not the inconsistencies but semantic concretizations in the present perfect tense. In order to appreciate the presence of the record, it will be required to have an inverted perspective. First, we should start from assigning an arbitrary set of semantic inconsistencies to the monetary economy, and then let each economic agent cope with bilateral transactions within the horizon of bilateral synchronization so as to meet the recorded semantic concretization registered in the present perfect tense. Each agent participates in bilateral transactions as observing the syntactic stipulation of fulfilling monetary flow continuity. Nonetheless, the presence of semantic inconsistencies makes each activity of syntactic integration precipitate further semantic inconsistencies on and on. Once semantic inconsistencies are assigned to the monetary economy, they can continue to survive since then. The judgement of whether the initial semantic inconsistencies as a working premise may be the right ones would entirely be up to how well the recorded semantic concretizations could be reproduced. The initial semantic inconsistencies to be assigned should be alternated by other choices until the faithful retracement of the recorded semantic concretizations becomes in sight.

Movement in progress upon semantic inconsistencies as retrieved from the record registered in the present perfect tense is markedly different from another movement stated fundamentally in the present tense. Mechanics, whether classical or quantal or whether non-relativistic or relativistic, is stated in the present tense as embodied in the equations of motion while being supplemented by the boundary conditions specified in the present perfect tense. Since it is stated in the present tense, mechanistic movement remains valid at any present moment, In addition, what is implied in mechanistic movement has already been completed and perfected in that it has already been supplemented and fixed by the boundary conditions specified in the present perfect tense. The future development following a mechanistic movement has already been completed and finished. The future sun eclipses have already been completed in the mechanistic model called celestial mechanics. There is no stressful dichotomy between syntactic integration and semantic concretization in mechanistic movement because no semantic inconsistencies are allowed to intervene in the latter.

In contrast, movement upon semantic inconsistencies is intrinsically irreversible in that the inconsistencies to be dealt with always come from the events already done. Retrieval of the movement upon semantic inconsistencies at the remote past does not require to refer to the record registered at the more recent past. It differs from mechanistic movement, in the latter of which the events in the record are taken to faithfully observe the movement whenever they have been registered. The unidirectionality of movement upon semantic inconsistencies now furnishes itself with the capacity of predicting or anticipating its own future development even to a limited extent. The monetary economy is just the case in point.

**5 Anticipation in the Monetary Economy**

Monetary flow equilibration as the activity of fulfilling monetary flow continuity at each economic agent is both syntactic and semantic. It is syntactic in that if preceding semantic inconsistencies are left unattended, the monetary economy would lose its organization. At the same time, monetary flow equilibration is semantic in that the syntactic integration necessarily leaves behind some inconsistencies in the semantic domain. Both the activities are certainly incommensurable because each belongs to a different category. Syntax refers to the linguistic structure, while semantics goes beyond the structure and reaches the empirical domain at large. Nonetheless, there is a perspective through which one can reduce semantic inconsistencies indirectly into a syntactic attribute. Insofar as one takes the activity of syntactic integration to be kept intact whatever semantic inconsistencies may come up, those semantic inconsistencies can be reduced to organizational disturbances acted upon the part of syntactic integration. Syntactic organization always is to come with organizational disturbances.

An organization constantly disturbing its own organization, that is imputed to the interplay between syntactic integration and semantic inconsistencies, now turns out to possess a unique dynamic characteristic. Organizational disturbances are flexible enough to cover a wide variety of structural variances, while the activity of organization is reliable enough to maintain its structural integrity. If a means of identifying the structure of an organization becomes available, the most likely one would be the one that can remain most robust against the accompanying organizational disturbances (Matsuno, 1978b). In fact, such a means of structural identification is undoubtedly available insofar as one appeals to the syntactic structure of a language.

We can thus envision a possible strategy of predicting the dynamic behavior of the monetary economy as reminding us of having recourse to retrieving the movement in the present progressive tense from the record registered in the present perfect tense. There have already been a few attempts for retrieving the dynamic movement of the monetary economy in progress from the record of the national financial statistics complied by the monetary authority of Japan (Ono, 1991; Hirano and Paton, 1999). A significant aspect to be noted in this context is that the actual organization of the monetary economy is the one remaining most robust against organizational disturbances of its own from within. This observation can easily be employed for making a prediction of the dynamic behavior of the monetary economy.

Just for the sake of argument, suppose we are interested in the development of these figures specifying the activities of the monetary economy such as the outstanding of the bank notes issued by the central bank, the discount rate at the central bank, the amount of securities issued by the government, the interest rate applied to the governmental securities, among others. The development of these figures can be represented as a trajectory in the multi-dimensional phase space, whose dimensions depend upon how many different figures we would like to choose to specify the dynamics. Then, we prepare an arbitrary set of possible future trajectories in the multi-dimensional phase space, each of which starts from a common initial point in the phase space. Once such an arbitrary set of trajectories has been prepared, each trajectory may be seen as the one in the record already registered in the present perfect tense. Each record now enables us to retrieve the corresponding movement in progress with recourse to the underlying semantic inconsistencies. We can also estimate its robustness against the organizational disturbances from within. The most likely future trajectory can thus be identified as the one that would remain most robust against the disturbances from within.

## 6 Concluding Remarks

Complex organization appearing in the empirical domain, whatever it may be, has kept attracting our attention for long by now. The issue of complex organization will be more sharply focused if one tries to predict or anticipate its future course even to an extremely limited extent. The most difficult part in the effort of prediction and anticipation rests upon the fact that the linguistic means coping with complex organization belongs to the syntactic domain while the object of interest belongs to the semantic or empirical domain. The problem is how one can reach organizational changes in the syntax from experiences in the empirical world. One clue for breaking the impasse which we have tried is to shed a legitimate light on the present progressive mode already available in the linguistic institution, since it is the present progressive tense that can connect both empirical activities and their linguistic representation most directly. What is unique to the present progressive tense is the two different roles it holds. The present progressive is syntactic in its occurrence, while it is semantic in its effect. The present progressive tense carries with itself both the capacities of organization and organizational disturbances.

Once it is legitimately recognized that any organization comes with the accompanying organizational disturbances, one can readily utilize this observation of an organization coming hand in hand with its own disturbances for the purpose of making a prediction or anticipation of the organization. The most likely organization must be the one remaining most robust as being subject to disturbances from within.